\documentclass[pra,twocolumn,a4paper,showpacs]{revtex4}
\usepackage{graphicx}
\usepackage{palatino}
\usepackage{bbm}
\usepackage{amsmath,amssymb}
\usepackage{latexsym}

\newcommand{\be}{\begin{equation}}
\newcommand{\ee}{\end{equation}}
\newcommand{\bea}{\begin{eqnarray}}
\newcommand{\eea}{\end{eqnarray}}
\newcommand{\Id}[1] {\int \! \! {\rm d}^3 #1}

\newcommand{\ve} {{\bf e}}
\newcommand{\vm} {{\bf m}}
\newcommand{\vj} {{\bf j}}
\newcommand{\vA} {{\bf A}}
\newcommand{\vB} {{\bf B}}
\newcommand{\bfsigma}{\mbox{\boldmath$\sigma$}}
\newcommand{\nn} {\nonumber}

\def\v#1{\mbox{\boldmath $#1$}}

\begin{document}
\title{Optimized Effective Potential Method in Current-Spin Density 
Functional Theory}
\author{S. Pittalis, S. Kurth, N. Helbig and E.K.U. Gross}
\affiliation{Institut f\"ur Theoretische Physik, Freie Universit\"at Berlin, 
Arnimallee 14, D-14195 Berlin, Germany}
\begin{abstract}
Current-spin density functional theory (CSDFT) provides a framework to 
describe interacting many-electron systems in a magnetic field which couples 
to both spin- and orbital-degrees of freedom. Unlike in usual (spin-) density 
functional theory, approximations to the exchange-correlation energy based 
on the model of the uniform electron gas face problems in practical 
applications. In this work, explicitly orbital-dependent functionals are used 
and a generalization of the Optimized Effective Potential (OEP) method to the 
CSDFT framework is presented. A simplifying approximation to the resulting 
integral equations for the exchange-correlation potentials is suggested. 
A detailed analysis of these equations is carried out for the case of  
open-shell atoms and numerical results are given using the exact-exchange 
energy functional. For zero external magnetic field, a small systematic 
lowering of the total energy for current-carrying states is observed due to 
the inclusion of the current in the Kohn-Sham scheme. For states without 
current, CSDFT results coincide with those of spin density functional theory. 
\end{abstract}

\maketitle

\section{Introduction}

Density functional theory (DFT) \cite{HohenbergKohn:64,KohnSham:65} is the 
standard tool to calculate the electronic structure of interacting 
many-electron systems. The original theorems of DFT have been successively 
extended to cover a wide range of different physical situations. One of the 
most important extensions is the spin-DFT (SDFT) formalism 
\cite{BarthHedin:72} which allows to describe systems with magnetic ground 
states in arbitrary external magnetic fields. In SDFT, however, 
the magnetic field only couples to the electronic spin while the coupling 
to the orbital degrees of freedom is not taken into account. To include also 
the coupling to the orbital currents, one has to resort to the 
current-spin-density functional theory (CSDFT) of Vignale and Rasolt
\cite{VignaleRasolt:87,VignaleRasolt:88}. 

Conceptually, DFT, SDFT, and CSDFT are very similar: they all map the system 
of interacting electrons onto a system of non-interacting particles in some 
effective fields. In the case of DFT, this auxiliary system yields the same 
electron density as the interacting one while in SDFT the magnetization 
densities of the two systems coincide as well. In CSDFT, also the paramagnetic 
current density of the 
auxiliary system is equal to the paramagnetic current density of the true, 
interacting system. In all three formalisms the energy of the interacting 
system is written as a functional of the corresponding densities and the value 
for the ground state energy is obtained by  minimizing this functional with 
respect to the densities. 

While all three flavors of DFT are exact in principle, in practice they all 
require an approximation for the exchange-correlation (xc) energy (which is a 
piece of the total energy) as a functional of the respective densities. Both 
in DFT and in SDFT, approximations based on the uniform electron gas such as 
the local (spin-) density aprroximation (L(S)DA) are surprisingly successful 
and they also form a good starting point for the construction of more 
sophisticated functionals such as generalized-gradient approximations (GGA's). 

In CSDFT, however, the situation is different. While it is possible to 
construct LDA-type approximations along similar ideas as in (S)DFT, these 
functionals are awkward to use in practical calculations 
\cite{SkudlarskiVignale:93} for a clear physical reason: when a uniform 
electron gas is exposed to an external magnetic field, Landau levels form 
and, for given magnetic field, the xc energy density exhibits derivative 
discontinuities as function of the density whenever a new Landau level 
starts to be filled. If this xc energy density is then used as the main 
ingredient in the construction of an LDA, these discontionuities 
show up in the corresponding xc potentials at those points 
in space where the local densities coincide with the densities of the uniform 
gas for which these discontinuities occur. This makes practical caluclations 
extremely difficult. One solution to this problem is to smoothly interpolate 
the xc energy density between the limits of weak and strong magnetic fields 
\cite{SaarikoskiRaesaenenSiljamaekiHarjuPuskaNieminen:03,WensauerRoessler:04} 
but this interpolation then misses the physics in the exchange-correlation 
energy arising from the Landau levels. 

The problem described above is entirely due to the construction of the LDA 
in CSDFT. Since the appearance of Landau levels is intrinsically an orbital 
effect, the use of explicitly orbital-dependent approximations to the xc 
energy functional offers a promising alternative which we explore in the 
present work. In (S)DFT, orbital functionals have attracted increasing 
interest in recent years \cite{GraboKreibichKurthGross:00} since these 
approximations offer a cure for notorious problems like, e.g., the 
self-interaction error of local and semilocal functionals. The calculation of 
the xc potential corresponding to an orbital-dependent xc energy functional is 
technically somewhat more involved than for explicitly density-dependent 
approximations and can be achieved with the so-called Optimized Effective 
Potential (OEP) method \cite{TalmanShadwick:76}. Here we present the extension 
of the OEP method to the CSDFT formalism \cite{HelbigKurthPittalisGross} and 
derive a set of coupled OEP integral equations for the corresponding 
exchange-correlation potentials. We then introduce a simplifying approximation 
in the spirit of Krieger, Li, and Iafrate (KLI) 
\cite{KriegerLiIafrate:92,KriegerLiIafrate:92-2} which transforms the integral 
equations into a set of algebraic equations which can be solved more easily 
in practical calculations. We present the resulting equations for fully 
non-collinear effective magnetic fields. This generalizes earlier work on 
a non-collinear implementation of orbital functionals in SDFT 
\cite{SharmaDewhurstDraxlHelbigKurthGrossShallcrossNordstroem}. 
It is also similar in spirit to a recent work \cite{RohraGoerling:06} which, 
however, uses a much larger set of densities. 

In a previous paper \cite{HelbigKurthPittalisGross} we have solved the OEP 
equations of CSDFT for two-dimensional quantum dots in external magnetic 
fields. In the present work we study the case of atoms at zero external 
magnetic field using the exact-exchange energy functional. In particular, we 
are interested in studying open-shell atoms which generally have 
degenerate ground states. It is well-known 
\cite{ZieglerRaukBaerends:77,BaerendsBranchadellSodupe:97} that in (S)DFT 
common approximations for the exchange-correlation energy do not lead to 
the same total energy for the different ground-state configurations. Here we 
investigate this problem in the framework of CSDFT because some of these 
ground states actually carry a non-zero current and one might hope that 
CSDFT is better suited to describe the degeneracies. The motivation to study 
simple atomic systems also aims at a better understanding of the differences 
between SDFT and CSDFT. The same orbital functional may perform quite 
differently in the three different schemes.

The present paper is organized as follows: in Section \ref{csdft} we give the 
fundamental equations of CSDFT before we present the derivation of the 
non-collinear OEP- and KLI-equations in Section \ref{oep}. This is followed in 
Section \ref{atoms} by an analysis of the properties of the KLI-potentials for 
open-shell atoms at zero external magnetic field. Numerical results 
are presented in Section \ref{numerics} before we draw our conclusions. 

\section{Current-Spin Density Functional Theory}
\label{csdft}

In this Section we briefly describe the formalism of current-spin density 
functional theory as originally suggested by Vignale and Rasolt 
\cite{VignaleRasolt:87,VignaleRasolt:88}. With this extension of the 
original DFT \cite{HohenbergKohn:64,KohnSham:65} formulation it becomes 
possible to study interacting many-electron systems in external magnetic 
fields. The CSDFT approach also goes beyond the widely used spin-DFT 
formalism \cite{BarthHedin:72} in the sense that while in SDFT the magnetic 
field only couples to the spin degrees of freedom, in CSDFT it also couples 
to the orbital degrees of freedom through the vector potential. 

The Hamiltonian of interacting electrons in an external electrostatic 
potential $v_0({\bf r})$ and an external magnetic field 
$\vB_0({\bf r}) = \nabla \times \vA_0({\bf r})$ is given by (atomic 
units are used throughout)
\bea
\hat{H} &=& \hat{T} + \hat{W} + \Id{r} \; \hat{n}({\bf r}) v_0({\bf r}) 
-\Id{r} \; \hat{\vm}({\bf r}) \vB_0({\bf r}) \nn \\
&& + \frac{1}{c} \Id{r} \; \hat{\vj}_p({\bf r}) \vA_0({\bf r}) + 
\frac{1}{2 c^2} \Id{r} \; \hat{n}({\bf r}) \vA_0^2({\bf r})  ,
\eea
where $\hat{T}$ and $\hat{W}$ are the operators of the kinetic energy and 
the electron-electron interaction, respectively. The operators for the 
density, the magnetization density, and the paramagnetic current density, 
are given by 
\be
\hat{n}({\bf r}) = \hat{\Psi}^{\dagger}({\bf r}) \hat{\Psi}({\bf r}) \; ,
\ee
\be
\hat{\vm}({\bf r}) = - \mu_B \hat{\Psi}^{\dagger}({\bf r}) {\bfsigma} 
\hat{\Psi}({\bf r}) \; ,
\ee
and
\be
\hat{\vj}_p({\bf r}) = \frac{1}{2 i} \left( \hat{\Psi}^{\dagger}({\bf r}) \nabla 
\hat{\Psi}({\bf r}) - (\nabla \hat{\Psi}^{\dagger}({\bf r})) 
\hat{\Psi}({\bf r}) \right) \; ,
\ee
respectively. Here we have defined field operators $\hat{\Psi}^{\dagger}({\bf r})
= (\hat{\psi}_{\uparrow}^{\dagger}({\bf r}), \hat{\psi}_{\downarrow}^{\dagger}({\bf r}))$ 
for two-component spinors, i.e., the formulation is not restricted to 
collinear magnetism with all the spins aligned in a single direction. 
$\bfsigma$ is the vector of Pauli matrices and 
$\mu_B= 1/(2 c)$ is the Bohr magneton.

Following Vignale and Rasolt \cite{VignaleRasolt:87,VignaleRasolt:88}, the 
ground state energy can be written as a functional of the three densities as 
\bea\label{tot-en1}
\lefteqn{
E[n,\vm,\vj_p] = F[n,\vm,\vj_p] + \Id{r} \; v_0({\bf r}) n({\bf r})} 
\nonumber \\
&& - \Id{r}\; 
\vm({\bf r}) \vB_0({\bf r}) + \frac{1}{c} \Id{r}\; \vj_p({\bf r}) 
\vA_0({\bf r}) \nonumber \\
&& + \frac{1}{2 c^2} \Id{r} \; n({\bf r}) \vA_0^2({\bf r}) \;,
\label{etot}
\eea
where $F[n,\vm,\vj_p]$ is a universal functional of the densities $n$, $\vm$, 
and $\vj_p$ in the sense that it is independent of the external fields $v_0$, 
$\vB_0$, and $\vA_0$. It may be decomposed in the usual way as 
\be
F[n,\vm,\vj_p] = T_s[n,\vm,\vj_p] + U[n] + E_{xc}[n,\vm,\vj_p] \;, 
\label{hk-func}
\ee
where $T_s[n,\vm,\vj_p]$ is the kinetic energy functional for non-interacting 
electrons, 
\be
U[n] = \frac{1}{2} \Id{r} \Id{r}' \; \frac{n({\bf r}) n({\bf r}')}{| {\bf r} - {\bf r}'|}
\ee
is the classical electrostatic energy and $E_{xc}[n,\vm,\vj_p]$ is the 
exchange-correlation energy. 

Using Eq.~(\ref{hk-func}) and minimizing the total energy (\ref{etot}) leads 
to the effective single-particle Kohn-Sham (KS) equations of CSDFT which 
read
\bea
\left[\frac{1}{2}
\left(\!-i\nabla+\frac{1}{c} \vA_s({\bf r})\right)^2+v_s({\bf r})+ 
\mu_B \v\sigma \vB_s({\bf r})\right]\Phi_i({\bf r}) \nonumber \\
= \epsilon_i \Phi_i({\bf r}) \;,
\label{ks-cdft}
\eea
where $\Phi_i({\bf r})$ are two-component, single-particle Pauli spinors. 
The effective potentials are given by
\be
v_s({\bf r}) = v_0({\bf r})+v_H({\bf r})+ v_{xc}({\bf r})
+\frac{1}{2c^2}\left[\vA_0^2({\bf r})-\vA_s^2({\bf r})\right] \; ,
\ee
\be
\vB_s({\bf r}) = \vB_0({\bf r})+\vB_{xc}({\bf r}) \; ,
\ee
and 
\be
\vA_s({\bf r}) = \vA_0({\bf r})+\vA_{xc}({\bf r}) \; ,
\ee
where the Hartree potential is given by 
\be
v_H({\bf r}) = \Id{r'} \frac{n({\bf r}')}{| {\bf r} - {\bf r}'|} \; .
\ee
The exchange-correlation potentials are functional deri\-vatives of the 
exchange-corre\-lation energy $E_{xc}$ with respect to the corresponding 
conjugate densities 
\be
v_{xc}({\bf r}) = \frac{\delta E_{xc}[n,\vm,\vj_{p}]}{\delta n({\bf r})} \;,
\label{vxc-c1}
\ee
\be
\vB_{xc}({\bf r}) = - \frac{\delta E_{xc}[n,\vm,\vj_{p}]}{\delta \vm({\bf r})} \;,
\label{bxc-c1}
\ee
\be
\vA_{xc}({\bf r}) = c \; \frac{\delta E_{xc}[n,\vm,\vj_{p}]}{\delta \vj_p({\bf r})} 
\; .
\label{axc-c1}
\ee
The effective potentials are such that the ground-state densities 
of the Kohn-Sham system reproduce those of the interacting system. The 
particle density can then be computed by 
\be
n({\bf r}) = \sum_{i=1}^{N} \Phi_i^{\dagger}({\bf r}) \Phi_i({\bf r}) \; ,
\label{density}
\ee
the magnetization density by
\be
\vm({\bf r}) = - \mu_B \sum_{i=1}^{N} \Phi_i^{\dagger}({\bf r}) \v\sigma 
\Phi_i({\bf r})  \; ,
\ee
and the paramagnetic current density by 
\be 
\vj_{p}({\bf r})= \frac{1}{2 i} \sum_{i=1}^{N}  
\left[
\Phi_{i}^{\dagger}({\bf r}) \nabla \Phi_{i}({\bf r}) 
- \left( \nabla \Phi_{i}^{\dagger}({\bf r}) \right) \Phi_{i}({\bf r}) 
\right] \; ,
\label{mc}
\ee
where the sums in Eqs.~(\ref{density}) - (\ref{mc}) run over the occupied 
Kohn-Sham spinor orbitals.

As usual in DFT, the exact form of the exchange-correlation energy functional 
$E_{xc}[n,\vm,\vj_{p}]$ is unknown and needs to be approximated in practice. 
In the present work, we focus on a class of approximate functionals which also 
has attracted increasing interest in SDFT in recent years. This is the class 
of functionals which explicitly depend on the Kohn-Sham orbitals and are 
therefore only implicit functionals of the densities. In our context these 
functionals are appealing for two reasons: first, they are constructed 
without any reference to the uniform electron gas and second, they are 
ideally suited to describe the appearance of Landau levels which in itself 
may be viewed as an orbital effect. 

\section{Optimized Effective Potential method in CSDFT}
\label{oep}
The calculation of the exchange-correlation potential for orbital-dependent 
functionals in ordinary (S)DFT is done in the framework of the Optimized 
Effective Potential (OEP) method 
\cite{SharpHorton:53,TalmanShadwick:76,GraboKreibichKurthGross:00}. This 
method derives its name from the fact that it yields that local potential 
whose orbitals minimize the total energy functional of the interacting 
system. This optimized potential is obtained as a solution of the so-called 
OEP integral equation. 

Recently, the OEP equations have been derived for the non-collinear 
formulation of SDFT 
\cite{SharmaDewhurstDraxlHelbigKurthGrossShallcrossNordstroem}. 
Here we derive the OEP integral equations for the exchange-correlation 
potentials in CSDFT \cite{HelbigKurthPittalisGross} by calculating the 
functional derivatives of
$E_{xc}$ with respect to the three effective potentials $v_s$, $\vB_s$, and
$\vA_s$. Making use of the correspondence both between Kohn-Sham spinors and 
ground-state densities, as well as between Kohn-Sham spinors and effective 
potentials, these functional derivatives can be computed in two different 
ways by using the chain rule, i.e., 

\begin{widetext}
\be
\label{oep-int1}
\frac{\delta E_{xc}}{\delta v_s({\bf r})}  
= \Id{r'} \biggl[
v_{xc}({\bf r}')\frac{\delta n({\bf r}')}{\delta v_s({\bf r})}
+\frac{1}{c}\vA_{xc}({\bf r}')\frac{\delta \vj_p({\bf r}')}{\delta 
v_s({\bf r})}
-\vB_{xc}({\bf r}')\frac{\delta\vm({\bf r}')}{\delta v_s({\bf r})}\biggl]
=\sum_{i=1}^{N}\Id{r'} \biggl[
\frac{\delta E_{xc}}{\delta\Phi_i({\bf r}')}\frac{\delta
\Phi_i({\bf r}')}{\delta v_s({\bf r})}
+ h.c. \biggl] \; ,
\ee
\be
\label{oep-int2}
\frac{\delta E_{xc}}{\delta \vB_s({\bf r})}
= \Id{r'} \biggl[
v_{xc}({\bf r}')\frac{\delta n({\bf r}')}{\delta \vB_s({\bf r})}
+\frac{1}{c}\vA_{xc}({\bf r}')\frac{\delta \vj_p({\bf r}')}{\delta 
\vB_s({\bf r})}
-\vB_{xc}({\bf r}')\frac{\delta\vm({\bf r}')}{\delta \vB_s({\bf r})}\biggl]
= \sum_{i=1}^{N}\Id{r'} \biggl[ 
\frac{\delta E_{xc}}{\delta\Phi_i({\bf r}')}\frac{\delta
\Phi_i({\bf r}')}{\delta \vB_s({\bf r})}
+ h.c \biggl] \; ,
\ee
\be
\label{oep-int3}
\frac{\delta E_{xc}}{\delta \vA_s({\bf r})}
=\Id{r'} \biggl[
v_{xc}({\bf r}')\frac{\delta n({\bf r}')}{\delta \vA_s({\bf r})}
+\frac{1}{c}\vA_{xc}({\bf r}')\frac{\delta \vj_p({\bf r}')}{\delta 
\vA_s({\bf r})}
-\vB_{xc}({\bf r}')\frac{\delta\vm({\bf r}')}{\delta \vA_s({\bf r})}\biggl]
=\sum_{i=1}^{N}\Id{r'} \biggl[
\frac{\delta E_{xc}}{\delta\Phi_i({\bf r}')}\frac{\delta
\Phi_i({\bf r}')}{\delta \vA_s({\bf r})}
+h.c.\biggl].
\ee
\end{widetext}
Eqs.~(\ref{oep-int1}) to (\ref{oep-int3}) constitute a system of coupled 
integral equations for the unknown exchange-correlation potentials. 
For simplicity, we have assumed that the approximation for $E_{xc}$ to be 
used depends only on the occupied spinor orbitals such as, e.g., the 
exact-exchange functional 
\bea
\lefteqn{
E_x^{EXX}[\{ \Phi_i\}] } \nonumber \\
&=& - \frac{1}{2} \sum_{i,j}^{{\rm occ}} \Id{r} \Id{r'} \; 
\frac{\Phi_i^{\dagger}({\bf r}) \Phi_j({\bf r}) \Phi_j^{\dagger}({\bf r}') 
\Phi_i({\bf r}')}{| {\bf r} - {\bf r}' |}
\label{exx}
\eea
which we use in our numerical implementation. 

For any approximation 
of $E_{xc}$ given explicitly in terms of the spinor orbitals, the 
functional derivatives of $E_{xc}$ with respect to these spinors can 
be evaluated easily. The other functional derivatives in 
Eqs.~(\ref{oep-int1}) - (\ref{oep-int3}) may be computed exactly 
from first-order perturbation theory by considering variations of the 
Kohn-Sham spinors due to a perturbation $\delta \hat{H}_s({\bf r}')$ of the 
Kohn-Sham Hamiltonian. To first order in the perturbation these variations are
\be
\delta \Phi_i({\bf r}) = 
\sum_{\stackrel{j=1}{j \neq i}}^{\infty}
\frac{\Phi_j({\bf r})}{\epsilon_i-\epsilon_j} \int d^3r' \Phi_j^{\dagger}({\bf r}') 
\delta \hat{H}_s({\bf r}')\Phi_i({\bf r}') \; ,
\label{delta-Phi}
\ee
where for simplicity we have assumed a non-degenerate spectrum. For arbitrary 
variations $\delta v_s({\bf r})$, $\delta \vB_s({\bf r})$, and $\delta \vA_s({\bf r})$ in 
the three effective potentials, the perturbation $\delta H_s({\bf r})$ is given by 
\bea
\lefteqn{
\delta \hat{H}_s({\bf r}) = \delta v_s({\bf r}) + \frac{1}{2ic} \nabla \delta 
\vA_s({\bf r}) } \nn \\ 
&& + \frac{1}{ic} \delta \vA_s({\bf r})\nabla + \frac{1}{c^2} 
\vA_s({\bf r})\delta \vA_s({\bf r})
+ \mu_B \v\sigma \delta \vB_s({\bf r})\;.
\eea
Insertion into Eq.~(\ref{delta-Phi}) allows us to identify the functional 
derivatives of the spinors with respect to the effective potentials as 
\be
\frac{\delta \Phi_i({\bf r})}{\delta v_s({\bf r}')} = 
\sum_{ \stackrel{j=1}{j \neq i}}^{\infty} \frac{\Phi_j({\bf r})}
{\epsilon_i-\epsilon_j} \left[\Phi_j^{\dagger}({\bf r}')\Phi_i({\bf r}')\right]\;,
\label{rfv}
\ee
\be
\frac{\delta \Phi_i({\bf r})}{\delta \vB_s({\bf r}')} = 
\mu_B\sum_{\stackrel{j=1}{j \neq i}}^{\infty}
\frac{\Phi_j({\bf r})}{\epsilon_i-\epsilon_j}
\left[\Phi_j^{\dagger}({\bf r}') \v\sigma \Phi_i({\bf r}')\right]\;,
\label{rfb}
\ee
and
\begin{widetext}
\be
\frac{\delta \Phi_i({\bf r})}{\delta \vA_s({\bf r}')} = 
\sum_{\stackrel{j=1}{j \neq i}}^{\infty}
\frac{\Phi_j({\bf r})}{\epsilon_i-\epsilon_j}
\left\{\frac{1}{2ic}\left[\Phi_j^{\dagger}({\bf r}') \nabla' \Phi_i({\bf r}') 
- \left(\nabla' \Phi_j^{\dagger}({\bf r}') \right) \Phi_i({\bf r}') \right] + \frac{1}{c^2} 
\vA_s({\bf r}) \Phi_j^{\dagger}({\bf r}')\Phi_i({\bf r}') \right\}\;.
\label{rfa}
\ee
\end{widetext}
From Eqs.~(\ref{rfv}) - (\ref{rfa}) one can compute the static 
response functions, i.e., the functional derivatives of the densities with 
respect to the effective potentials. Inserting everything into 
Eqs.~(\ref{oep-int1}) - (\ref{oep-int3}) one can then write the OEP integral 
equations in a very compact form as 
\be
\label{oep1}
\sum_{i=1}^N \Phi_i^{\dagger}({\bf r})\Psi_i({\bf r}) + h.c. = 0, \; 
\ee
\be
\label{oep2}
-\mu_B\sum_{i=1}^N \Phi_i^{\dagger}({\bf r})\v\sigma\Psi_i({\bf r}) + h.c. 
= 0 \;,
\ee
and
\be
\label{oep3}
\frac{1}{2i}\sum_{i=1}^N 
\left\{\Phi_i^{\dagger}({\bf r})\nabla\Psi_i({\bf r})
-\left[\nabla\Phi_i^{\dagger}({\bf r})\right]\Psi_i({\bf r})\right\} + h.c. = 0 \;,
\ee
where we have defined the so-called orbital shifts 
\cite{GraboKreibichKurthGross:00,KuemmelPerdew:03}
\be
\Psi_i({\bf r})=\sum_{\stackrel{j=1}{j \neq i}}^{\infty}
\frac{\Phi_j({\bf r})D_{ij}}{\epsilon_i-\epsilon_j},
\label{orb-shift}
\ee
with
\bea 
\lefteqn{
D_{ij}= \Id{r'} \biggl(
v_{xc}({\bf r}')\Phi_j^{\dagger}({\bf r}')\Phi_i({\bf r}') } \nn \\
&& +\frac{1}{2 i c}\vA_{xc}({\bf r})
\left[\Phi_j^{\dagger}({\bf r}')\nabla'\Phi_i({\bf r}')
-\left(\nabla'\Phi_j^{\dagger}({\bf r}')\right)\Phi_i({\bf r}')\right] \nn \\
&&+\mu_B\vB_{xc}({\bf r}')\Phi_j^{\dagger}({\bf r}')\v\sigma
\Phi_i({\bf r}') -\Phi_j^{\dagger}({\bf r}')\frac{\delta E_{xc}}{\delta 
\Phi_i^{\dagger}({\bf r}')} \biggl).
\eea
The name ``orbital shifts'' (\ref{orb-shift}) derives from their structure 
as a first-order shift from the unperturbed Kohn-Sham orbital $\Phi_i$ under a 
perturbation whose matrix elements are given by $D_{ij}$. The OEP equations 
(\ref{oep1})-(\ref{oep3}) have a very simple interpretation: they merely say 
that the densities do not change under this perturbation. Keeping in mind 
that the Kohn-Sham system already yields the exact densities, this statement 
is actually quite obvious. 

As already mentioned, the OEP equations are a set of coupled integral 
equations for the exchange-correlation potentials. In this work we do not 
attempt a full solution of these equations but rather suggest a simplifying 
approximation \cite{SharpHorton:53} in the spirit of Krieger, Li, and Iafrate 
(KLI) \cite{KriegerLiIafrate:92,KriegerLiIafrate:92-2} who introduced the same 
approximation in the usual OEP method of SDFT. In this KLI approximation 
the orbital shifts are approximated by replacing the energy denominator by 
some constant, i.e., 
\be
\Psi_i({\bf r}) \approx \frac{1}{\Delta\epsilon} \left( 
\sum_{j=1}^{\infty} \Phi_j({\bf r})D_{ij} - \Phi_i({\bf r})D_{ii} \right) \; .
\label{orb-shift-2}
\ee
Inserting this approximation into the OEP equations and applying the 
completeness relation for the Kohn-Sham spinors one obtains after some 
algebra a set of algebraic equations for the exchange-correlation 
potentials which can conveniently be written as 
\be
{\cal D}({\bf r}) {\cal V}_{xc}({\bf r}) = {\cal R}({\bf r}) \; .
\label{kli-matrix}
\ee
Here, we have defined the 7-component vector ${\cal V}_{xc}({\bf r})$ as 
\be
{\cal V}_{xc}^T({\bf r}) = \left(v_{xc}({\bf r}),\vB_{xc}^T({\bf r}),\frac{1}{c} 
\vA_{xc}^T({\bf r})\right)
\ee
and the $7 \times7$ matrix ${\cal D}({\bf r})$ has the structure
\bea
\nonumber
{\cal D}({\bf r}) = \left( 
\begin{array}{ccc}
n({\bf r})     & - \vm^T({\bf r}) & \vj_p^T({\bf r})\\
- \vm({\bf r}) & \mu_B^2 n({\bf r}) {\mathbbm 1} & {\cal J}({\bf r}) \\
\vj_p({\bf r})  & {\cal J}^T({\bf r}) & {\cal N}({\bf r}) 
\end{array}
\right), 
\eea
where ${\mathbbm 1}$ is the $3 \times3$ unit matrix. The matrix elements of 
the $3 \times 3$ matrices ${\cal J}$ and ${\cal N}$ are defined by
\be
{\cal J}_{\alpha \beta}({\bf r}) = - \frac{\mu_B}{2 i} \sum_{i=1}^N \!\left(\!
\Phi_i^{\dagger}({\bf r}) \sigma_{\alpha} \frac{\partial \Phi_i({\bf r})}
{\partial r_{\beta} } - \frac{\partial \Phi_i^{\dagger}({\bf r}) }
{\partial r_{\beta}} \sigma_{\alpha} \Phi_i({\bf r})\!\right)\!\!,
\ee
and 
\bea\nonumber
{\cal N}_{\alpha \beta}({\bf r}) &=& \frac{1}{2} \sum_{i=1}^N \left( 
\frac{\partial \Phi_i^{\dagger}({\bf r})}{\partial r_{\alpha}} 
\frac{\partial \Phi_i({\bf r})}{\partial r_{\beta}} + 
\frac{\partial \Phi_i^{\dagger}({\bf r})}{\partial r_{\beta}}
\frac{\partial \Phi_i({\bf r})}{\partial r_{\alpha}} \right)\\
&& - \frac{1}{4 n({\bf r})} \frac{\partial n({\bf r})}{\partial r_{\alpha}} 
\frac{\partial n({\bf r})}{\partial r_{\beta}}\;, 
\eea
where $\alpha=1,2,3$ corresponds to the cartesian coordinates $x$, $y$, and 
$z$, respectively. 
The seven components of the vector ${\cal R}({\bf r})$ on the right-hand side of 
Eq.~(\ref{kli-matrix}) are given by
\be
\label{eq-r1}
{\cal R}_1({\bf r}) = \frac{1}{2} \sum_{i=1}^N \left( \Phi_i^{\dagger}({\bf r}) 
\frac{\delta E_{xc}}{\delta \Phi_i^{\dagger}({\bf r})} + n_i({\bf r}) D_{ii} + h.c. \right), 
\ee
\bea
\lefteqn{
{\cal R}_{1+\alpha}({\bf r}) = \frac{1}{2} \sum_{i=1}^N } \nn \\
&& \Bigg( - \mu_B \Phi_i^{\dagger}({\bf r}) 
\sigma_{\alpha} \frac{\delta E_{xc}} {\delta \Phi_i^{\dagger}({\bf r})} 
+ m_{i,\alpha}({\bf r}) D_{ii} + h.c. \Bigg)  \; ,
\label{eq-r2-4}
\eea
\bea
\lefteqn{
{\cal R}_{4+\alpha}({\bf r}) = \frac{1}{2} \sum_{k=1}^N \Bigg( 
\frac{1}{2 i} \Phi_k^{\dagger}({\bf r}) \frac{\partial}{\partial r_{\alpha}} 
\frac{\delta E_{xc}}{\delta \Phi_k^{\dagger}({\bf r})} } \nn \\
&& -  \frac{1}{2 i} 
\frac{\partial \Phi_k^{\dagger}({\bf r})}{\partial r_{\alpha}} 
\frac{\delta E_{xc}}{\delta \Phi_k^{\dagger}({\bf r})} + j_{p,k,\alpha}({\bf r}) D_{kk} + h.c. \Bigg) 
\label{eq-r5-7}
\eea
with the density $n_i({\bf r})$, magnetization density 
$\vm_i({\bf r})$, and paramagnetic current density $\vj_{p,i}({\bf r})$ of the single 
orbital $\Phi_i({\bf r})$. 
It is worth mentioning that in order to arrive at this result we used the 
identity  \cite{VignaleRasolt:88}
\be
\nabla \biggl( n({\bf r})\vA_{xc}({\bf r}) \biggr) = 0
\label{gauge-prop}
\ee
which follows directly from gauge invariance of the exchange-correlation 
energy. 

The KLI equations (\ref{kli-matrix}) can be solved by iteration: 
start with an intial guess for the potentials to compute the orbitals 
and the right-hand side of Eq.~(\ref{kli-matrix}), then solve this equation 
for the new potentials and iterate until self-consistency is achieved. 

In a different work \cite{HelbigKurthPittalisGross} we have solved the 
KLI equations for a two-dimensional quantum dot in an external magnetic field. 
In the present work we use our CSDFT-OEP formalism to study open-shell atoms 
in zero external magnetic field. In the next Section we discuss some 
further assumptions we employed in our implementation and deduce some 
analytic results for the KLI potentials. 

\section{Open-shell atoms at zero magnetic field}
\label{atoms}

We want to employ our CSDFT-OEP formalism to study open-shell atoms. From the 
point of view of CSDFT this is interesting since some states out of the 
multiplet of degenerate ground states have a non-vanishing current density 
while others do not carry a current. 

In the limit of zero external magnetic field, the Kohn-Sham equation 
(\ref{ks-cdft}) takes the form 
\bea
\Bigg( -\frac{\nabla^2}{2} + v_0({\bf r}) + v_H({\bf r}) + v_{xc}({\bf r})
+ \frac{1}{2ic} \left(\nabla \vA_{xc}({\bf r}) \right) \nn \\
+ \frac{1}{i c} 
\vA_{xc}\nabla + \mu_{B} 
{\bfsigma} \vB_{xc}({\bf r}) \Bigg) \Phi_{i}({\bf r}) 
= \epsilon_{i} \Phi_{i}({\bf r})  .
\label{ksc}
\eea
We further employ the collinear approximation assuming that the Kohn-Sham 
spinors decompose into spin-up ($\sigma=+1$) and spin-down ($\sigma=-1$) 
orbitals, i.e., $\Phi_i({\bf r}) = (\varphi_{i, \sigma=+1}({\bf r}),0)$ or 
$\Phi({\bf r})=(0,\varphi({\bf r})_{i, \sigma=-1})$. As a result the magnetization 
density is parallel to the $z$-direction, $\vm({\bf r}) = (0,0,m({\bf r}))$.
In addition, we assume cylindrical symmetry for both the densities and
the corresponding conjugate potentials (that is they do not depend on the 
azimuthal angle $\phi$). As a consequence the magnetic quantum number
is a good quantum number for the single-particle orbitals which then take 
the form
\be 
\varphi_{i m \sigma}({\bf r}) = g_{i \sigma}(r,\theta) \exp(im\phi ) 
\chi(\sigma) \;,
\label{ksso}
\ee
where we used radial coordinates and $m$ is the magnetic quantum number 
(not to be confused with $m({\bf r})$, the $z$-component of the magnetization 
density). $\sigma$ is the spin index and $\chi(\sigma)$ is the eigenfunction 
of the $z$-component of the spin operator. 
In the collinear approximation,  $\vB_{xc}({\bf r})= (0,0,B_{xc}({\bf r}))$ is 
parallel to the $z$-axis while $\vA_{xc}({\bf r})= A_{xc}({\bf r}) \ve_{\phi}$ 
where $\ve_{\phi}$ is the unit vector in $\phi$-direction. As an 
additional consequence of the cylindrical symmetry of our problem we have 
$\nabla \vA_{xc}({\bf r}) = 0$. 

We restrict ourselves to ground states whose densities can be reproduced by a 
single Slater determinant. For example, for the boron atom one configuration 
has all three up-electrons and the two down-electrons in states with magnetic 
quantum number $m=0$ while in another configuration one of the up-electrons 
occupies an $m=1$ state with the other occupations unchanged. In this way
current-carrying and zero-current states can be considered. The resulting 
current only has a component in the $\phi$-direction, 
$\vj_p({\bf r}) = j_{p}({\bf r}) \ve_{\phi}$. We may then rewrite Eq.~(\ref{ksc}) as 
\bea
\Bigg( -\frac{\nabla^2}{2} + v_{0}({\bf r}) + v_{H}({\bf r}) + v_{xc}({\bf r}) 
+ \frac{1}{c} \frac{m}{r \sin \theta }A_{xc}({\bf r})  \nn \\
  + \mu_{B} \sigma B_{xc}({\bf r}) \Bigg) \varphi_{i m \sigma}({\bf r}) 
= \epsilon_{i m \sigma} \varphi_{i m \sigma}({\bf r}) \; . 
\label{ksca}
\eea

In the following we discuss a number of typical cases: 
For atomic closed-shell configurations, where the 
density is spherical and both the magnetization density and the paramagnetic 
current density vanish identically, both $A_{xc}$ and $B_{xc}$ vanish 
identically. Obviously, in this situation CSDFT reduces to the original 
density-only DFT. 

For ground-state configurations where only orbitals with $m = 0$ are 
occupied, the correct value for $j_{p}({\bf r})$ - which is zero at any point 
in space - is trivially obtained already within the SDFT scheme. Therefore  
we expect that $v_{xc}({\bf r}) = v_{xc}^{SDFT}({\bf r})$, 
$B_{xc}({\bf r})=B_{xc}^{SDFT}({\bf r})$, and $A_{xc}({\bf r}) = 0$. Actually, any other 
choice of $A_{xc}({\bf r})$ would not make any difference for the ground state
densities. In a way this may be regarded as a simple manifestation of the
non-uniqueness of the CSDFT potentials pointed out in 
Ref.~\cite{CapelleVignale:02}. 

As a third case we consider ground-state configurations with a half-filled 
shell as in, e.g., the nitrogen atom. Again, SDFT already gives the 
correct values of the total densities. Therefore, we again expect that 
$v_{xc}({\bf r}) = v_{xc}^{SDFT}({\bf r})$, $B_{xc}({\bf r})=B_{xc}^{SDFT}({\bf r})$ and 
$A_{xc}({\bf r}) = 0$. 

Ground-state configurations carrying a non-vanish\-ing paramagnetic current
are the most interesting ones in our context. At zero external magnetic 
field, this situation only arises for open-shell atoms away from half-filling. 
Indeed, it is for these states that we expect $A_{xc}({\bf r}) \ne 0$ 
as well as $v_{xc}({\bf r}) \ne v_{xc}^{SDFT}({\bf r})$, $B_{xc}({\bf r}) 
\ne B_{xc}^{SDFT}({\bf r})$.

In the following we analyze the KLI equations for the above cases 
in order to confirm these expectations. We remind the reader 
that in our derivation of the OEP equations we assumed that $E_{xc}$ depends  
only on the occupied orbitals. Moreover, we also assume that 
\be
\frac{\delta E_{xc}}{\delta \varphi_{i m \sigma}({\bf r})} \sim \exp(-im \phi)
\ee
holds. Both of these assumptions are correct for the exact-exchange functional 
which we consider in our numerical implementation. 

Under these assumptions, the first two KLI-equations (for $\sigma = \pm 1$) are
\be
v_{xc,\sigma}({\bf r}) + \frac{1}{c} \frac{j_{p,\sigma}({\bf r})}{n_{\sigma}({\bf r})} 
A_{xc}({\bf r}) = w^{(1)}_{xc,\sigma}({\bf r}) + w^{(2)}_{xc,\sigma}({\bf r}) \; ,
\label{kli1}
\ee
where we have defined the spin-dependent scalar potential 
\be
v_{xc,\sigma}({\bf r}) = v_{xc}({\bf r}) + \mu_{B} \sigma B_{xc}({\bf r}) \; .
\ee
The terms on the right-hand side of Eq.~(\ref{kli1}) are given by
\be
w^{(1)}_{xc,\sigma}({\bf r})=\frac{1}{n_{\sigma}({\bf r})} \sum_{i, m}^{{\rm occ}} 
n_{i m \sigma}({\bf r}) u_{xc, i m \sigma}({\bf r})
\ee
and
\be
w^{(2)}_{xc,\sigma}({\bf r}) = \frac{1}{n_{\sigma}({\bf r})} \sum_{i, m}^{{\rm occ}} 
n_{i m \sigma}({\bf r}) d_{xc,i m \sigma}
\ee
with
\be
u_{xc,i m \sigma}({\bf r}) = \frac{1}{\varphi^{*}_{i m \sigma}({\bf r})} 
\frac{\delta E_{xc}}{\delta \varphi_{i m \sigma}({\bf r})}
\ee
and
\bea
d_{xc,i m \sigma} &=& \Id{r} \; n_{i m \sigma}({\bf r}) \left( 
v_{xc,\sigma}({\bf r}) - u_{xc,i m \sigma}({\bf r}) \right) \nn \\
&& + \frac{1}{c} \Id{r} \; j_{p,i m \sigma}({\bf r}) A_{xc}({\bf r}) \;.
\eea
Here $n_{i m \sigma}({\bf r})$ and $j_{p,i m \sigma}({\bf r})$ are the density and the 
paramagnetic current density of the single orbital $\varphi_{i m \sigma}({\bf r})$ 
which, for our symmetry, are related by 
\be
j_{p,i m \sigma}({\bf r}) = m \; \frac{n_{i m \sigma}({\bf r})}{r \sin \theta} \;.
\ee

The third KLI equation reads
\bea
\left( \sum_{\sigma} j_{p,\sigma}({\bf r})v_{xc,\sigma}({\bf r}) \right)  + 
\frac{1}{c} N({\bf r}) A_{xc}({\bf r}) = \nn \\
\sum_{\sigma} \left( \tilde{w}^{(1)}_{xc,\sigma}({\bf r}) + 
\tilde{w}^{(2)}_{xc,\sigma}({\bf r}) \right)
\label{kli2}
\eea
with 
\be
N({\bf r}) = \sum_{\sigma}  \sum_{i, m}^{{\rm occ}}  
\frac{j_{p,i m \sigma}^2({\bf r})}{n_{i m \sigma}({\bf r})} \;,
\ee
\be
\tilde{w}^{(1)}_{xc,\sigma}({\bf r}) = \sum_{i, m}^{{\rm occ}} 
j_{p,i m \sigma}({\bf r}) u_{xc,i m \sigma}({\bf r}) \;,
\ee
and
\be
\tilde{w}^{(2)}_{xc,\sigma}({\bf r}) = \sum_{i, m}^{{\rm occ}} 
j_{p,i m \sigma}({\bf r}) d_{xc,i m \sigma} \; .
\ee
It is interesting to note that $v_{xc,\sigma}({\bf r})$ and $A_{xc}({\bf r})$ couple 
to each other only if at least one of the $j_{p,\sigma}({\bf r})$ is non-vanishing. 

At this point, we again consider open-shell configurations for which all 
occupied orbitals have $m=0$. Then $N({\bf r})$ vanishes and the KLI 
equation (\ref{kli1}) reduces to the KLI equation of SDFT while 
Eq.~(\ref{kli2}) becomes a trivial identity. As a consequence, 
$v_{xc,\sigma}({\bf r}) = v_{xc,\sigma}^{SDFT}({\bf r})$ and $A_{xc}({\bf r})$ is 
undetermined. As discussed above, $A_{xc}({\bf r})$  does not affect any of the 
ground state densities and we fix it as $A_{xc}({\bf r})=0$. 

Next we consider configurations with a half-filled shell. We assume that we
have already solved the SDFT KLI equations and use the resulting orbitals 
and potentials plus the initial guess $A_{xc}({\bf r})=0$ as a start for 
the iterative solution of the CSDFT KLI equations. We substitute this initial 
guess into Eqs.~(\ref{kli1}) and (\ref{kli2}) to compute the new potentials.
The occupied orbitals of SDFT either have $m=0$ or they come in 
pairs with $m$ and $-m$. This leads to the same contributions to 
$u_{xc,\sigma,i}({\bf r})$ for orbitals in the same shell while for the 
paramagnetic current they contribute with equal magnitude but opposite sign. 
Hence, the KLI equations become
\be
v^{new}_{xc,\sigma}({\bf r}) = 
w^{(1)}_{xc,\sigma}({\bf r}) + w^{(2)}_{xc,\sigma}({\bf r}) = 
v_{xc,\sigma}^{SDFT}({\bf r})
\label{kli1_2}
\ee
and
\be
\frac{1}{c} N({\bf r})A^{new}_{xc}({\bf r}) = 0 \Rightarrow A^{new}_{xc}({\bf r}) = 0\;.
\label{kli2_2}
\ee
This shows that the SDFT solution along with $A_{xc}({\bf r})=0$ is also a CSDFT 
solution. We also tested numerically that the solution $A_{xc}({\bf r})=0$ is 
stable against (not necessarily small) perturbations of the initial guess. 

Finally, we consider the most interesting case of ground-state configurations 
with a paramagnetic current different from zero. For these configurations we 
expect $A_{xc}({\bf r}) \ne 0$. Solution of the third KLI-equation (\ref{kli2}) 
with respect to  $A_{xc}({\bf r})$ yields
\bea
\lefteqn{
A_{xc}({\bf r}) = }\nn \\
&& c \; \frac{\sum_{\sigma} \left( \tilde{w}^{(1)}_{xc,\sigma}({\bf r}) + 
\tilde{w}^{(2)}_{xc,\sigma}({\bf r}) -  j_{p,\sigma}({\bf r})v_{xc,\sigma}({\bf r})
\right)}{N({\bf r})} \; .
\label{kli2bis}
\eea
In this equation, the denominator increases for increasing number of 
electrons. The numerator also typically increases when more orbitals are 
occupied but, due to large cancellations for contributions arising from 
orbitals with opposite values of $m$, it increases with a slower rate than 
the denominator. As a consequence, we expect larger exchange-correlation 
vector potentials $A_{xc}({\bf r})$ for atoms in the first row than for atoms 
in the second row (but the same column) of the periodic table.

In the remainder of this Section we discuss the asymptotic behavior of 
the exchange-correlation potentials and the vector potential. 

We start by {\em assuming} that, for finite systems, the exchange-correlation 
potentials in the asymptotic region far away from the system behave as 
\be
v_{xc,\sigma}({\bf r})  \stackrel{r \to \infty}{\longrightarrow} - \frac{1}{r}
\label{vxc-bound}
\ee
and
\be
\lim_{r \to \infty} A_{xc}({\bf r}) = 0 \; .
\label{axc-bound}
\ee
Eq.~(\ref{vxc-bound}) certainly is a reasonable assumption: it is the 
well-known asymptotic behavior for $v_{xc,\sigma}$ of SDFT which we assume to 
be unchanged in CSDFT. Eq.~(\ref{axc-bound}) then ensures that the term 
proportional to $A_{xc}({\bf r})/r$ in the Kohn-Sham equation (\ref{ksca}) decays 
faster than $v_{xc,\sigma}({\bf r})$ asymptotically. At this stage, 
Eq.~(\ref{axc-bound}) may be viewed as a working assumption in order to be 
able to proceed further. Below we show that it is consistent with the 
solution of the KLI equation.

Under this assumption we can deduce 
\cite{KreibichKurthGraboGross:99,GraboKreibichKurthGross:00} the asymptotic 
behavior of the atomic orbitals from the Kohn-Sham equation (\ref{ksca}) as 
\be
\lim_{r \to \infty} \varphi_{i m \sigma}({\bf r}) = r^{-1 + 1/\beta_{i m \sigma}} 
\exp(- \beta_{i m \sigma} r) \; ,
\ee
where $\beta_{i m \sigma} = \sqrt{-2 \epsilon_{i m \sigma}}$. This implies 
that the spin density is dominated asymptotically by the highest occupied 
orbital of that spin. The same is true for the current density if the magnetic 
quantum number of this orbital is different from zero (as typically is the 
case for current-carrying states of open-shell atoms). 

In order to proceed further with our analysis we restrict ourselves to the 
exact-exchange functional of Eq.~(\ref{exx}). Then the KLI equation 
(\ref{kli1}) allows us to establish the following relation between 
$v_{xc,\sigma}$ and $A_{xc}$ in the asymptotic region 
\be
\lim_{r \to \infty} v_{xc,\sigma}({\bf r}) + \frac{1}{c} \frac{M_\sigma}{r \sin 
\theta} A_{xc}({\bf r}) \to - \frac{1}{r} + d_{xc, N_\sigma M_{\sigma} \sigma} \; ,
\label{lim1}
\ee
where we tacitly assumed that we are taking the limit away from a nodal plane 
of the highest occupied orbital of spin $\sigma$ 
\cite{GraboKreibichKurthGross:00,DellaSalaGoerling:01}. Here $N_\sigma$ is the 
orbital index of that orbital and $M_\sigma$ is its magnetic quantum number. 
Since we are working in the collinear approximation, the Kohn-Sham equations 
for the two spin channels are completely decoupled and we can choose a 
constant shift in $v_{xc, \sigma}$ such that
\be
d_{xc, N_\sigma M_\sigma \sigma} = 0 \; .
\ee
Eq.~(\ref{lim1}) together with Eq.~(\ref{vxc-bound}) then implies 
\be
\frac{M_{\sigma}}{r \sin \theta } A_{xc}({\bf r}) 
\stackrel{r \to \infty}{\longrightarrow} 0 
\label{lim3}
\ee
which is consistent with the assumption of Eq.~(\ref{axc-bound}). 

However, a closer inspection of the KLI equations (\ref{kli1}) and 
(\ref{kli2}) shows that they become linearly dependent in the asymptotic 
region and therefore do not have a unique solution. This again may 
be viewed as a manifestation of the non-uniqueness problem in CSDFT 
\cite{CapelleVignale:01, CapelleVignale:02}. In our numerical procedure 
to be described in the next Section we take a pragmatic approach to the 
problem of linearly dependent KLI equations and choose a 
solution with $A_{xc}({\bf r}) \to 0$ and a $v_{xc,\sigma}({\bf r})$ satisfying
Eq.~(\ref{vxc-bound}).

Before concluding this Section, we discuss some symmetry properties 
of the exchange-correlation vector potential and exchange-correlation 
magnetic field. By inspection of the two KLI-equations (\ref{kli1}) and 
(\ref{kli2}) it is clear that under the exchange of spin-up and spin-down 
electrons, $B_{xc}({\bf r})$ changes sign. Similarly, exchanging an electron from 
an orbital with magnetic quantum number $m$ to a previously unoccupied one 
with $-m$ leads to $A_{xc}({\bf r})$ changing sign. These transformations can be 
performed independently leading to the same total energy.

\section{Numerical results for open-shell atoms at zero external 
magnetic field}
\label{numerics}

In this Section we describe the numerical results for open-shell atoms 
obtained within the KLI approximation of CSDFT using the exact-exchange 
functional of Eq.~(\ref{exx}). In particular, we are interested in 
calculating total energies of current-carrying and zero-current states
in various ground-state configurations. In principle, the states of
the ground-state multiplet should be degenerate but in SDFT
zero-current states are always lower in energy than current-carrying
states. Since the current appears to be the quantity leading to these
spurious energy splittings, it is interesting to see if CSDFT (where
the current is one of the fundamental variables) can alleviate the
problem. Since the main difference between SDFT and CSDFT is the 
appearance of an exchange-correlation vector potential in the Kohn-Sham 
equation, we also present some results for the xc potentials in 
the different approaches. 

We have developed an atomic code for CSDFT and SDFT calculations in a 
basis set representation, assuming cylindrical symmetry of the
Kohn-Sham potentials and densities. As basis functions we use
Slater-type functions for the radial part multiplied with spherical 
harmonics for the angular part. For the Slater functions we employ the 
quadruple zeta basis sets (QZ4P) of 
Ref.~\cite{VeldeBickelhauptBaerendsFonsecaGisbergenSnijdersZiegler:2001}.
We have tested our code by computing the total energies of spherically 
symmetric atoms of the first and second row of the periodic table in 
exchange-only KLI approximation and compared with results from accurate, 
fully numerical codes available in the literature 
\cite{KriegerLiIafrate:92,GraboKreibichKurthGross:00,Engel:03}. 
Our code reproduces these total energies to within an average deviation of 
0.1 kcal/mol for the atoms in the first row and to within an
average deviation of 0.5 kcal/mol for atoms in the second row
of the periodic table. 

We performed self-consistent exchange-only calculations in the KLI 
approximation of CSDFT for open-shell atoms in current-carrying and 
zero-current configurations. The configurations are selected by 
specifying the number of occupied states for each value of the magnetic 
quantum number $m$. Once a choice has been made, the occupations remain 
unchanged during the self-consistency cycle. In all the cases we studied we 
were able to obtain self-consistent solutions for both zero-current and 
current-carrying states. 

As expected, for zero-current states we always obtain a self-consistent 
CSDFT solution with vanishing exchange vector potential, 
$A_x({\bf r})\equiv 0$. In fact, this solution, which is equivalent to the 
corresponding SDFT solution, always gives the lowest total energy. 

\begin{figure}[t]
\includegraphics[width=0.9\columnwidth]{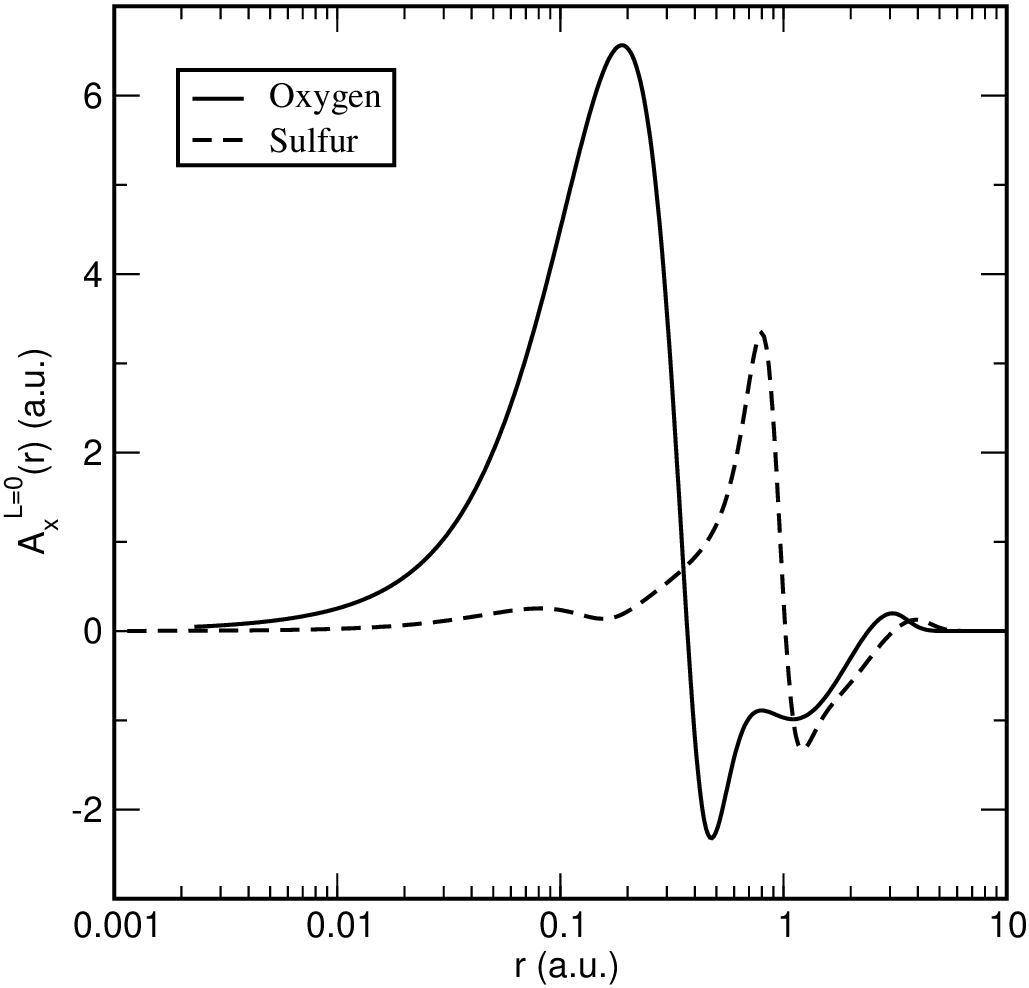}\\[3mm]
\caption{\label{fig1} Spherical component of the exchange vector potentials 
for current-carrying states of the oxygen and sulfur atoms.}
\end{figure}

For current-carrying states we always find a non-vanishing $A_x$. 
However, as a consequence of the 
linear dependence of the KLI equations (\ref{kli1}) and (\ref{kli2}) 
discussed in the previous Section, the exchange potential and vector potential 
are not uniquely determined. In fact, without additional numerical measures 
we obtain unphysical potentials which diverge asymptotically. This may lead to 
a wrong ordering of the occupied and unoccupied orbitals or even to 
convergence problems.
A numerically convenient scheme to use the KLI equations (\ref{kli1}) and 
(\ref{kli2}) during the self-consistency cycle is to first calculate 
$v_{xc,\sigma}({\bf r})$ from Eq.~(\ref{kli1}) with the $A_{xc}({\bf r})$ obtained in 
the previous iteration. In this step we impose the asymptotic behavior 
of $v_{xc,\sigma}({\bf r}) \stackrel{ r \to \infty}{\longrightarrow} -\frac{1}{r}$. 
 Then, with this updated $v_{xc,\sigma}({\bf r})$ we use 
Eq.~(\ref{kli2}) to obtain a new $A_{xc}({\bf r})$. In order to enforce the 
asymptotic limit $A_{xc}({\bf r}) \stackrel{ r \to \infty}{\longrightarrow} 0$
we add a small quantity $\delta$ to $N({\bf r})$ in Eq. (\ref{kli2bis}). 
Total energies and current densities are very insensitive to the choice of 
$\delta$: for the fixed value of $\delta = 10^{-4}$ a.u. which we use for all 
our calculations, total energies vary by an order of $10^{-2}$ kcal/mol or 
less if $\delta$ is varied by an order of magnitude around its chosen value.

In Fig.~\ref{fig1} we show the $L=0$ (i.e., spherical) component of an 
expansion of the exchange vector potentials in terms of Legendre polynomials, 
i.e., $A_x({\bf r})= \sum_{L=0}^{\infty} A_x^L(r) P_L(\cos \theta)$, for the oxygen 
and sulfur atoms in the current-carrying 
state. As we argued in the previous Section, the exchange vector potential 
of sulfur is smaller than the one for oxygen which also implies that the 
difference between SDFT and CSDFT total energies is smaller for the 
heavier atom. 

\begin{figure}[t]
\includegraphics[width=0.9\columnwidth]{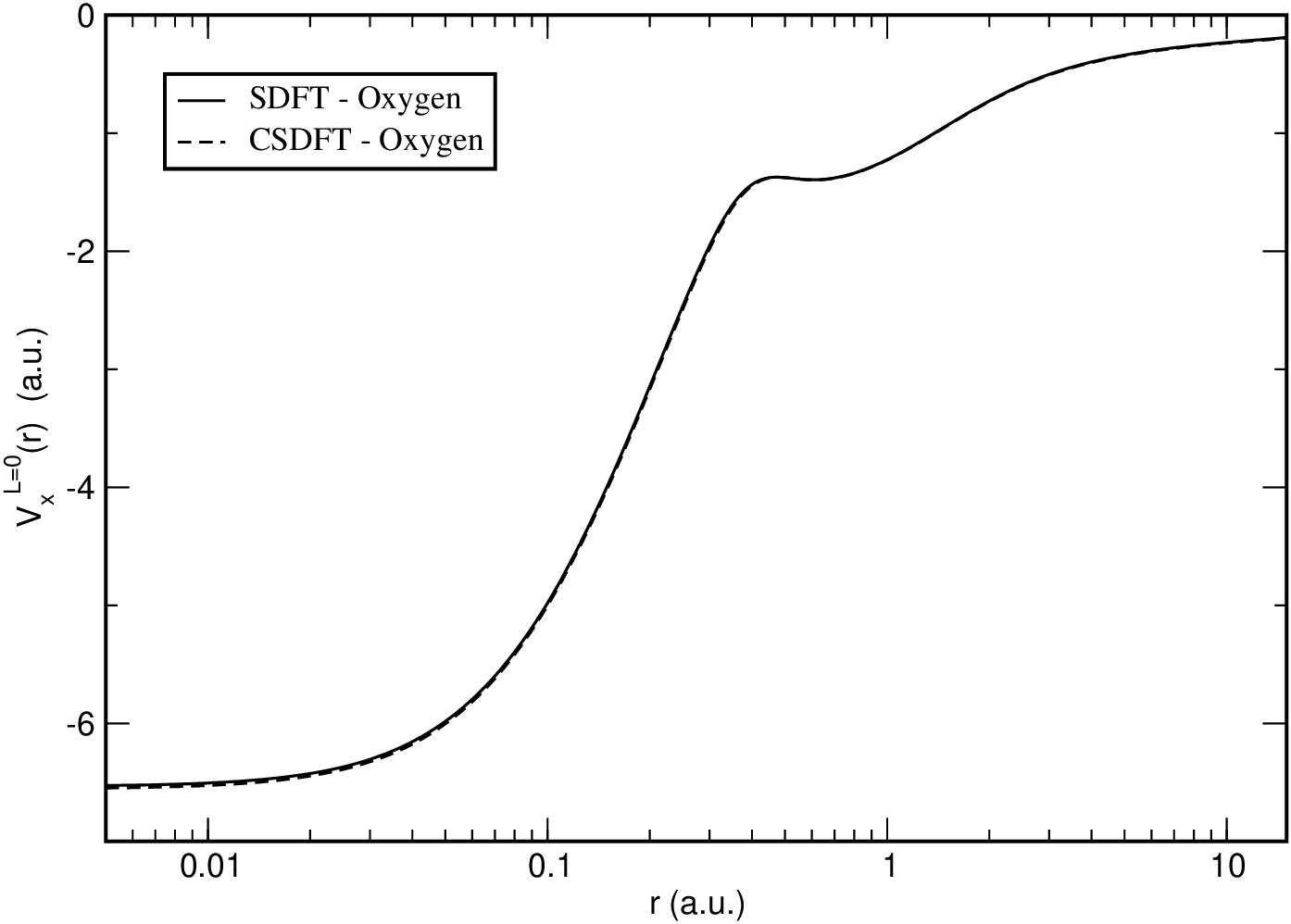}\\[3mm]
\caption{\label{fig2} Spherical component of the exchange scalar potentials 
for the current-carrying state of the oxygen atom computed in SDFT and CSDFT.}
\end{figure}

\begin{figure}[b]
\vspace*{3mm}
\includegraphics[width=0.9\columnwidth]{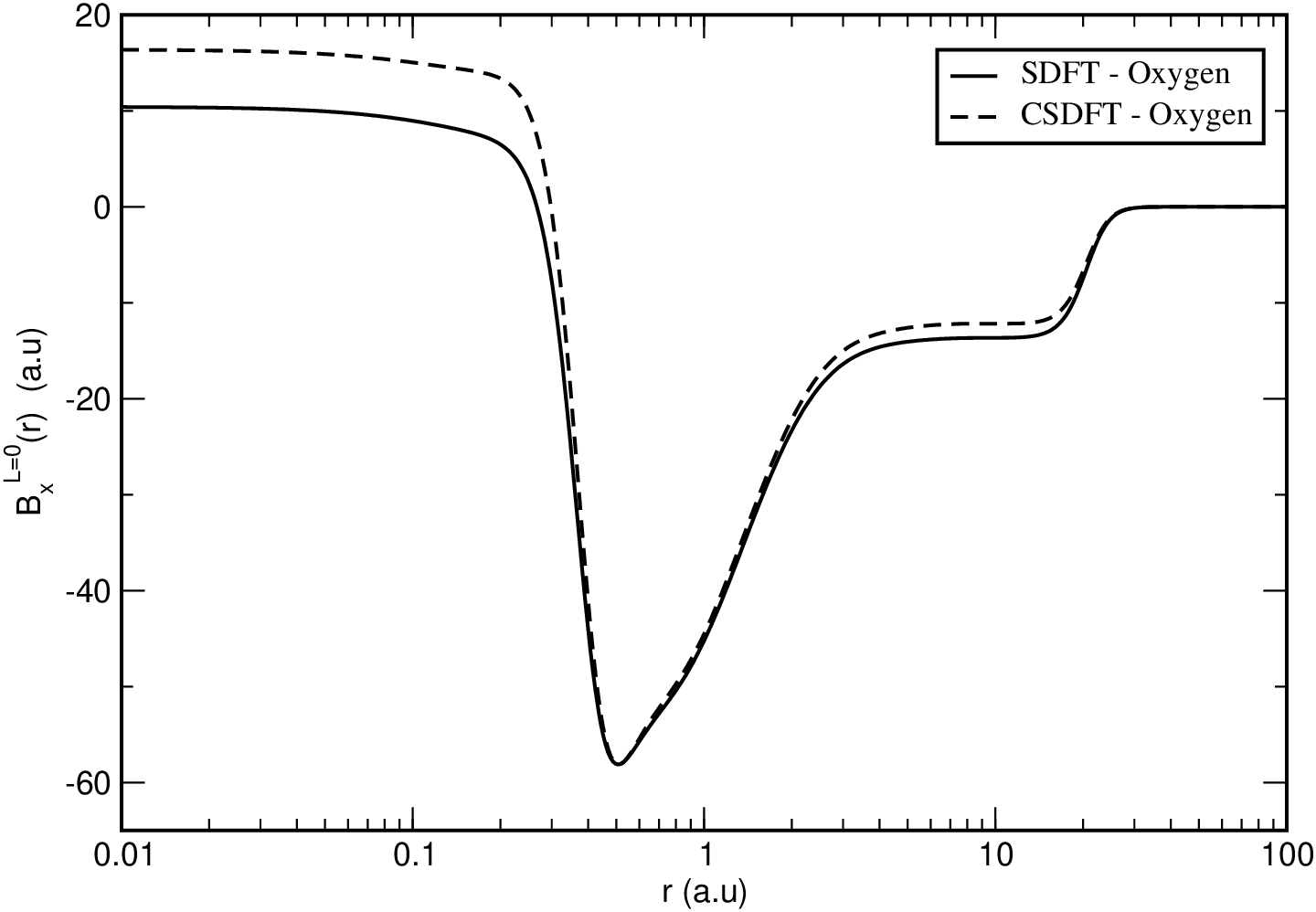}\\[3mm]
\caption{\label{fig3} Spherical component of the exchange magnetic field for 
the current-carrying state of the oxygen atom computed in SDFT and CSDFT.}
\end{figure}

In previous work 
\cite{Capelle:99,OrestesMarcassoCapelle:03,OrestesDaSilvaCapelle:05} it has 
been assumed that CSDFT and SDFT calculations lead to very similar results 
because the term associated with the vector potential is expected to be small. 
In order to verify this assumption, we  compare the self-consistent 
exchange potentials and magnetic fields for the current-carrying state of 
the oxygen atom for a SDFT and a CSDFT calculation. The spherical components 
of the exchange potentials in the two approaches are shown in Fig.~\ref{fig2}. 
The potentials are hardly distinguishable on the scale of the plot which 
confirms the initial assumption. 
If one looks at the corresponding exchange 
magnetic fields (Fig.~\ref{fig3}) one sees that the overall structure 
of the SDFT and the CSDFT results are quite similar. However, there are 
significant differences in magnitude close to the nucleus which can be 
expected to have a visible effect on the resulting chemical shifts 
\cite{BuehlKauppMalkinaMalkin:99}. This is also reflected by a substantial 
difference in the relative magnetization density 
$\zeta(0)= (n_{\uparrow}(0)-n_{\downarrow}(0))/(n_{\uparrow}(0)+n_{\downarrow}(0))$
at the nuclear position. For the current-carrying state of oxygen we obtain 
the value $\zeta(0)= -1.03 \times 10^{-3}$ in SDFT and 
$\zeta(0)= -1.16 \times 10^{-3}$ in CSDFT which amounts to a difference 
of approximately 13\%. While we are confident that the difference of SDFT and 
CSDFT values for $\zeta(0)$ is not a numerical artefact, the absolute numbers 
should be read with some caution. In order to estimate the accuracy of these 
numerically sensitive results we have also calculated the same quantity 
for the nitrogen atom and obtain $\zeta(0)= -1.77 \times 10^{-3}$. This value 
differs by approximately 9\% from the value $\zeta(0)=-1.62 \times 10^{-3}$ 
given in Table 10 of Ref.~\cite{GraboKreibichKurthGross:00} which was obtained 
with a fully numerical code for spherically symmetric effective potentials. 

Finally, we have calculated the spurious energy splittings between different 
configurations of open-shell atoms of the first two rows of the periodic 
table in SDFT and CSDFT. The results are collected in Table \ref{tab1}. 
As mentioned before, in all cases the splittings are positive, i.e., 
the zero-current states are always lowest in energy. The splittings themselves 
are systematically lower in CSDFT than in SDFT due to the additional 
variational degree of freedom in the former approach. The effect of 
including the vector potential is more significant for the lighter atoms 
while for the second row the CSDFT splittings are only less than 0.1 kcal/mol 
smaller than the ones obtained from SDFT. Although a CSDFT approach to the 
degeneracy problem appeared promising our results show only a small and 
insufficient improvement. This finding is somewhat at odds with recent 
suggestions to reduce the splittings by inclusion of the current density as 
a variable in the construction of approximate exchange-correlation 
functionals \cite{Becke:02,MaximoffErnzerhofScuseria:04,TaoPerdew:05}. These 
works, however, suggested orbital functionals in the framework of SDFT where 
the orbital dependence entered through the current density. No attempt 
was made to implement these functionals in a CSDFT framework. Finally, we 
point out that in a recent work \cite{PittalisKurthGross:06} we showed, that a 
pure DFT approach using the exact-exchange functional surprisingly leads to 
energy splittings which are almost an order of magnitude smaller than the 
ones obtained in SDFT.

\begin{table}[t]
\begin{tabular}{c|cc|c|cc}
\hline\hline 
Atom & $\Delta^{\mathrm{SDFT}}$ & $\Delta^{\mathrm{CSDFT}}$ & Atom & 
$\Delta^{\mathrm{SDFT}}$  &  $\Delta^{\mathrm{CSDFT}}$ \\ [0.5ex] 
\hline 
B  & 1.66 & 1.38 & Al & 1.68 & 1.58 \\ [0.5ex] 
C  & 1.58 & 1.34 & Si & 1.76 & 1.63 \\ [0.5ex] 
O  & 2.36 & 2.29 & S  & 3.04 & 3.01 \\ [0.5ex]
F  & 2.32 & 2.27 & Cl & 3.15 & 3.10 \\ [0.5ex]
\hline\hline
\end{tabular}
\caption{\label{tab1} 
Spurious energy splittings $\Delta = E(M= \pm 1)-E(M=0)$ (in kcal/mol) 
between current-carrying and zero-current states computed in SDFT and CSDFT.}
\end{table}

\section{Conclusions}

In this work, we have shown how one can use orbital-dependent functionals 
in the framework of current-spin density functional theory. We have derived 
the OEP integral equations which have to be solved to obtain the 
corresponding exchange-correlation potentials. We have simplified these 
integral equations in the spirit of the well-known KLI approximation. 

We have analyzed the KLI equations and the resulting potentials for open-shell 
atoms at zero external magnetic field and have also presented numerical 
results for these systems using the exact-exchange functional.
We have shown that CSDFT and SDFT are equivalent for the states with 
zero paramagnetic current. This equivalence breaks down for current-carrying 
states where total energies in CSDFT are lower than those of SDFT.  
 
We also verified that the CSDFT Kohn-Sham scheme leads to a reduction 
(compared to SDFT) of the spurious splittings between 
current-carrying and zero-current states although it is too small to recover 
the degeneracy between these states. 

The most important result of our study, however, is the fact that the 
problems of LDA-type current-density functionals derived from the uniform 
electron gas (such as unphysical discontinuities of the corresponding 
exchange-correlation potentials) never appear when using orbital dependent 
functionals. 

\section*{Acknowledgements}
We gratefully acknowledge financial support through the Deutsche Forschungsgemeinschaft
Priority Program 1145 "First-Principles Methods", 
through the EU Network of Excellence NANOQUANTA and through the EU Research and Training Network EXCITING.

\end{document}